\documentclass[aps,prb,twocolumn,showpacs,superscriptaddress]{revtex4-2}

\usepackage{graphicx}
\usepackage{amsmath}
\usepackage{multirow}
\usepackage{tabularx}
\usepackage{color}
\usepackage[colorlinks,linkcolor=blue,anchorcolor=blue,citecolor=blue]{hyperref}
\usepackage{xcolor}
\usepackage{ulem}

\begin{document}
	
	\title{Microscopic origin of magnetoferroelectricity in monolayer NiBr$_{2}$ and NiI$_{2}$}

	\author{Hui-Shi Yu}
	\affiliation{Center for Neutron Science and Technology, Guangdong Provincial Key Laboratory of Magnetoelectric Physics and Devices, State Key Laboratory of Optoelectronic Materials and Technologies, School of Physics, Sun Yat-Sen University, Guangzhou, 510275, China}
	\author{Xiao-Sheng Ni}
	\affiliation{Center for Neutron Science and Technology, Guangdong Provincial Key Laboratory of Magnetoelectric Physics and Devices, State Key Laboratory of Optoelectronic Materials and Technologies, School of Physics, Sun Yat-Sen University, Guangzhou, 510275, China}
        \affiliation{Peng Cheng Laboratory, Frontier Research Center, Shenzhen, China}
	\author{Dao-Xin Yao}
	\author{Kun Cao}
	\email{caok7@mail.sysu.edu.cn}
	\affiliation{Center for Neutron Science and Technology, Guangdong Provincial Key Laboratory of Magnetoelectric Physics and Devices, State Key Laboratory of Optoelectronic Materials and Technologies, School of Physics, Sun Yat-Sen University, Guangzhou, 510275, China}
	\begin{abstract}


We investigate the magnetoelectric properties of the monolayer NiX$_{2}$ (X = Br, I) through first-principles calculations. Our calculations predict that the NiBr$_{2}$ monolayer exhibits a cycloidal magnetic ground state. For the NiI$_{2}$ monolayer, a proper-screw helical magnetic ground state with modulation vector \(\boldsymbol{Q} = (q, 0, 0)\) is adopted, approximated based on experimental observations. The electric polarization in NiBr$_{2}$ shows a linear dependence on the spin-orbit coupling strength \(\lambda_{\text{SOC}}\), which can be adequately described by the generalized Katsura-Nagaosa-Balatsky (gKNB) model, considering contributions from up to the third nearest-neighbor spin pairs. In contrast, the electric polarization in NiI$_{2}$ exhibits a distinct dependence on \(q\) and \(\lambda_{\text{SOC}}\), which cannot be fully explained by the gKNB mechanism alone. To address this, the \(p\)-\(d\) hybridization mechanism is extended to NiI$_{2}$ to explain the observed behavior. The respective contributions from the \(p\)-\(d\) hybridization and the gKNB mechanism in NiI$_{2}$ are then quantitatively evaluated. Overall, our work elucidates the microscopic mechanisms underlying multiferroicity in NiBr$_{2}$ and NiI$_{2}$ monolayers, with the conclusions readily applicable to their bulk forms.
\end{abstract}
	
\maketitle	
\section{Introduction}
	
In recent years, there has been a renewed surge of interest in multiferroic materials that exhibit both magnetism and ferroelectricity~\cite{1,2}. One of the most promising applications is the ability to switch multiferroic domains in response to applied magnetic or electric fields, which can generate robust magnetoelectric responses for multifunctional electronic devices~\cite{3-1,3-2,3-3}. Among these, van der Waals (vdW) two-dimensional (2D) materials with multiferroic orders have attracted particular attention~\cite{matsukura2015control,huang2018electrical,jiang2018controlling}. Notably, vdW systems exhibiting type-II multiferroicity, where inversion symmetry is broken by magnetic ordering, have garnered significant interest. These systems not only offer new opportunities to explore novel mechanisms of magnetoelectric coupling in 2D~\cite{kurumaji2013magnetoelectric,kurumaji2020spiral,amoroso2021interplay}, but also enable the development of 2D-material-based nanoelectronic and spintronic applications~\cite{webster2018strain,burch2018magnetism,mak2019probing}.

Transition-metal dihalides MX$_{2}$ (M = transition metal cation, X = halogen anion) are a class of van der Waals materials that have attracted increasing interest due to their unique structures and remarkable properties. Among these, MnI$_{2}$, CoI$_{2}$, and NiX$_{2}$ (X = Br or I)~\cite{mcguire2017crystal,mak2019probing} have all exhibited type-II multiferroic behavior in their bulk forms~\cite{8-3, wang2016microscopic,7,kurumaji2013magnetoelectric}. Significant efforts have been dedicated to studying the properties of NiI$_{2}$ multilayers~\cite{ni}, demonstrating that NiI$_{2}$ remains multiferroic with a helical spin order even down to the bilayer~\cite{ju2021possible}. More recently, second harmonic generation (SHG) studies have indicated the presence of a multiferroic state in monolayer NiI$_{2}$ below a transition temperature of approximately 21 K~\cite{10-2}. However, there is ongoing debate regarding the interpretation of these experimental results, specifically whether the observed symmetry-breaking signal originates from magnetic orders or ferroelectric orders~\cite{wu2023layer,jiang2023dilemma}. To address this controversy, spin-polarized scanning tunneling microscopy (STM) experiments have been performed to determine the magnetic configuration of monolayer NiI$_{2}$, revealing a proper-screw spin-spiral state with a canted plane and a propagation vector \(\boldsymbol{Q}\) = (0.2203, 0, 0)~\cite{miao2309spin}. NiBr$_{2}$ has also attracted renewed interest as a multiferroic material~\cite{rai2019influence}, with substantial efforts devoted to investigating its few-layer forms~\cite{amoroso2020spontaneous,mushtaq2017nix,lu2019mechanical}. Experimental studies have indicated that monolayer NiBr$_{2}$ undergoes a noncollinear magnetic ordering below approximately 27 K, which is likely a helical magnetic state~\cite{amoroso2020spontaneous,2021noncollinear}.


Several theoretical studies have explored spin-induced ferroelectricity in NiX$_{2}$. Femega {\it et al.}~\cite{fumega2022microscopic} investigated the microscopic origin of multiferroic ordering in monolayer NiI$_{2}$ with a modulation vector \(\boldsymbol{Q} = (0.14, 0, 0)\), highlighting the coexistence of helical magnetic order and ligand spin-orbit coupling (SOC) as the primary driving forces behind the emergence of multiferroic behavior. 
The gKNB model is also applied to investigate monolayer NiI$_{2}$, obtaining results consistent with the corresponding numerical Monte Carlo (MC) simulations~\cite{10-2}. A more recent study suggests that Kitaev interactions play a crucial role in the formation of the proper-screw magnetic structure of bulk NiI$_{2}$~\cite{li2023realistic}. For monolayer NiBr$_{2}$, systematic first-principles calculations are conducted with experimental electric polarization reproduced within a spiral magnetic configuration~\cite{wu2023first}. However, a comprehensive investigation into the microscopic mechanisms of magnetoelectric coupling in monolayer NiX$_{2}$ remains lacking.


In this study, we investigate the magnetoelectric properties of the monolayer NiX$_{2}$ (X = Br, I) through first-principles calculations. Our calculations predict that the NiBr$_{2}$ monolayer exhibits a cycloidal magnetic ground state. For the NiI$_{2}$ monolayer, a proper-screw helical magnetic ground state with modulation vector \(\boldsymbol{Q} = (q, 0, 0)\) is adopted, approximated from experimental observations. The calculated electric polarization in NiBr$_{2}$ shows a linear dependence on the spin-orbit coupling strength \(\lambda_{\text{SOC}}\), which can be well explained by the gKNB model.
However, for the NiI$_{2}$ monolayer, the gkNB model alone cannot fully describe the distinct dependence of its electric polarization on \(q\) and \(\lambda_{\text{SOC}}\). This necessitates the introduction of an alternative theoretical interpretation.
We therefore apply the higher-order \( p \)-\( d \) hybridization mechanism and propose that the magnetoelectricity in the monolayer NiI$_2$ can be explained by the combination of the gKNB model and the \( p \)-\( d \) hybridization mechanism. Furthermore, the respective contributions from the \(p\)-\(d\) hybridization and the gKNB mechanism in NiI$_{2}$ are quantitatively extracted, utilizing the \(q\)-dependence and \(\lambda_{\text{SOC}}\)-dependence of the ferroelectric polarization. Overall, our work elucidates the microscopic mechanisms underlying multiferroicity in NiBr$_{2}$ and NiI$_{2}$ monolayers. The conclusions can be readily applied to their bulk forms.



\section{methods}
	
 Our first-principles calculations are performed using the Vienna \textit{ab initio} Simulation Package (VASP)~\cite{vasp-1,vasp-2}. The Perdew-Burke-Ernzerhof (PBE) functional with spin-polarized generalized gradient approximation (GGA) is employed for the exchange-correlation functional. A 20 $\times$ 20 $\times$ 4 k-point grid centered at $\Gamma$ ensures convergence using the Projector Augmented Wave (PAW) method with a 500 eV plane wave cutoff. A 15 \AA\ vacuum layer between adjacent monolayers along the z-direction prevents interlayer interactions. Spin-polarized GGA with onsite Coulomb interactions, $U$, is applied to Ni 3$d$ orbitals (GGA + $U$)~\cite{GGA+U}. For NiI$_{2}$ monolayers, $U = 4$ eV and $J = 1$ eV are chosen, yielding magnetic moments and transition temperatures consistent with experiments~\cite{ex1,ex2}. \(U_{\text{eff}} = 1\) eV~\cite{amoroso2020spontaneous} is used for NiBr$_{2}$ monolayer calculations. Structures are relaxed until the forces on each atom are below 1 meV/\AA. Heisenberg exchange interactions are calculated by fitting $J$ parameters to the DFT energies of randomly generated collinear magnetic configurations~\cite{12,13,14}. The Berry-phase method~\cite{berryphase} is employed to compute electric polarizations. We investigate the magnetic phase diagrams via a replica-exchange MC approach~\cite{cao2009first} based on the calculated magnetic exchange interactions. By applying Fourier analysis to the magnetic configurations obtained from MC simulations, the helical magnetic orders and corresponding magnetic propagation vectors $\boldsymbol{Q}$ are determined. For the gKNB calculations, the no-substitution method is used to determine the magnetoelectric coupling matrix $M$ for NiX$_{2}$ monolayers~\cite{xiang2011general}.

	\section{Results and discussions}
	
\begin{figure*}[t]  
	\centering  
	\includegraphics[width=1\textwidth]{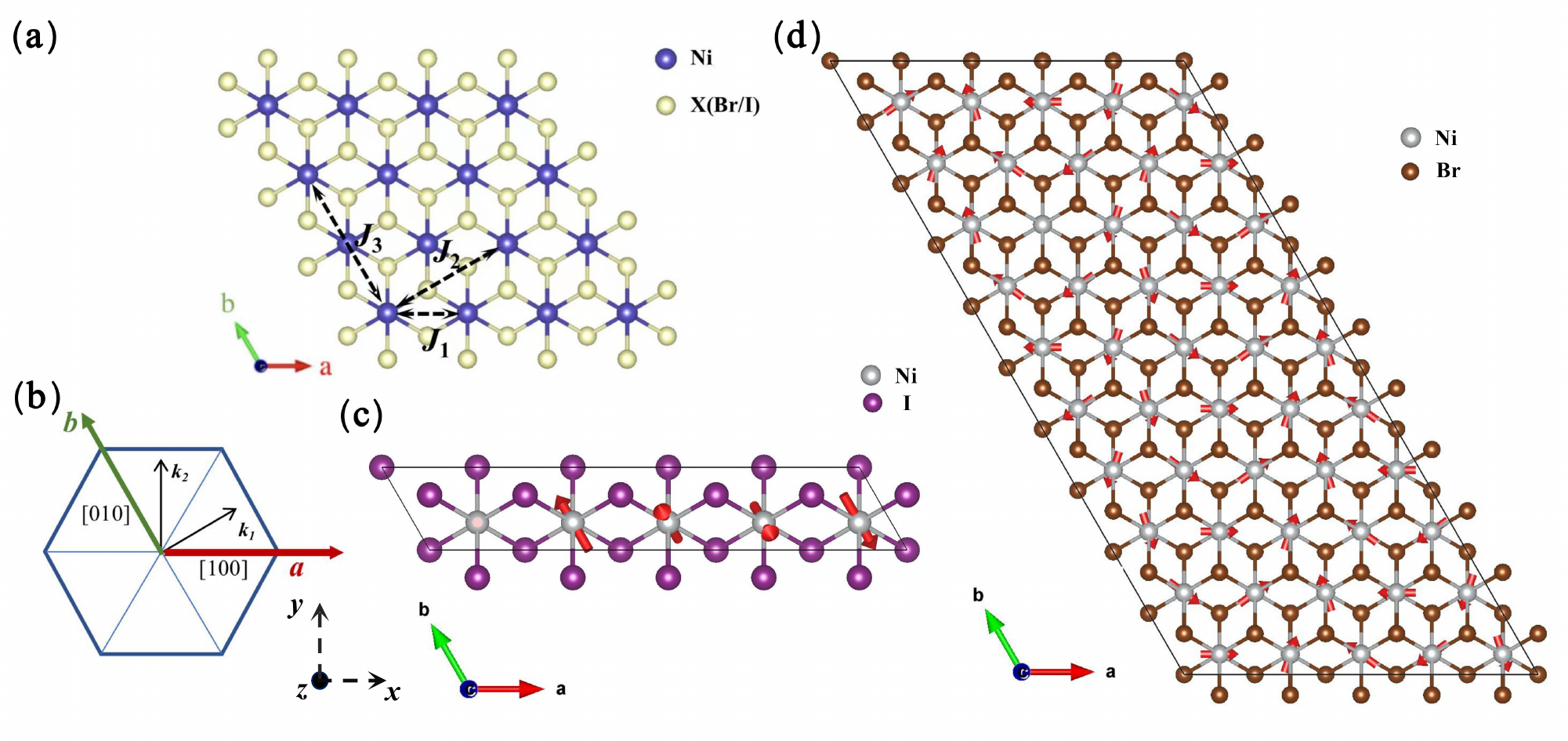}  
	\caption{(a) Top view of monolayer NiX$_{2}$. The exchange paths are shown in black dotted lines with double arrows.
(b)The in-plane lattice vectors, \( \mathbf{a} \) and \( \mathbf{b} \), along with the corresponding reciprocal lattice vectors, \( \mathbf{k}_1 \) and \( \mathbf{k}_2 \), and the Cartesian coordinate axes (\( x \), \( y \), and \(z \)) are presented.
(c) Representation of the HM order in monolayer NiI$_{2}$ with q=(0.2, 0, 0) in a 5 $\times$ 1 supercell. (d) The HM order with q=(0.2, -0.1, 0) in monolayer NiBr$_{2}$ shown in a  5 $\times$ 10 supercell.}
	\label{nix2}
\end{figure*}


Bulk NiX$_2$ crystallizes in the trigonal \textit{R$\overline{3}$m} (No. 166) space group, with experimental lattice constants of $a$ = $b$ = 4.46 \AA\ and c = 10.73 \AA\ for NiI$_{2}$~\cite{10-2,ju2021possible}, and  $a$ = $b$ = 3.65 \AA\ and c = 18.24 \AA\  for NiBr$_{2}$~\cite{mcguire2017crystal,day1976optical}. In Ref.~\cite{11}, Bulk NiI$_{2}$ enters into a helimagnetic (HM) multiferroic phase with a magnetic propagation vector of $\boldsymbol{Q}$ = (0.138, 0, 1.457) below a N\'{e}el temperature of 59.5 K. In bulk NiBr$_{2}$, the multiferroic state emerges below a transition temperature of 22.8 K~\cite{7}. All Ni atoms in bulk NiX$_2$ are 2+ valence with spin S = 1~\cite{lu2019mechanical,mcguire2017crystal}.  As the lattice reduces from bulk to the monolayer limit, the space group reduces to \textit{P$\overline{3}$m1} (No. 164)~\cite{an2022spin,lu2019mechanical}. More structural information of NiX$_2$ monolayer is provided in Table.~\ref{tab1}.

Our DFT calculations show that the NiBr$_2$ monolayer possesses smaller lattice constants compared to NiI$_2$. The Ni-Br bonds are also shorter than the Ni-I bonds, indicating a more compact lattice structure for the NiBr$_2$ monolayer. We consider a NiI$_2$ monolayer with an exact proper screw magnetic structure with an approximated modulation vector $\boldsymbol{Q}$ = (0.2, 0, 0),  as shown in Fig.~\ref{nix2}(c). The commensurate in-plane q-vector differs only by a $\sim$9$\%$ from the experimentally measured results of $\boldsymbol{Q}$ = (0.2203, 0, 0)~\cite{miao2309spin}.
For NiBr$_2$ monolayer, there is no conclusive magnetic structures from experimental measurements so far, therefore we first analyze its magnetic exchange interaction to unveil its magnetic configuration (see Fig.~\ref{nix2}).

\begin{table*}[t!]
		\caption{ \label{tab1} Structural data for monolayer NiX$_{2}$ compounds}
		\begin{ruledtabular}
			\begin{tabular}{cccccccccccc}
				& Compound & Structure Type & Lattice Constants &  $ d_{Ni-X} $ & $T_{C}$  \\
				\hline  \\[-1.0ex]
				& NiI$_{2}$ &  \textit{P$\overline{3}$m1} (No.164)~\cite{an2022spin} & $a$ = $b$ = 3.98 \AA\cite{an2022spin}\ & 2.76 \AA\ & 21.0 K~\cite{10-2} \\
				& NiBr$_{2}$ & \textit{P$\overline{3}$m1} (No.164)~\cite{lu2019mechanical} &  $a$ = $b$ = 3.69 \AA\cite{lu2019mechanical} & 2.56 \AA\ & 22.8 K (bulk)~\cite{7}  \\
			\end{tabular}
		\end{ruledtabular}
	\end{table*}


\subsection{Magnetic ground state of NiBr$_2$ monolayer}

 To model the exchange couplings of NiBr$_{2}$ monolayer, we use a spin Hamiltonian of the following form:
	\begin{equation}
		\begin{aligned}
			\textit{H}_{spin} = \sum_{ij}J_{ij}\textbf{S}_i\ \cdot\ \textbf{S}_j-\sum_{i}(\textbf{D}\ \cdot\ \textbf{S}_i)^{2}
		\end{aligned}\label{eq:1}
	\end{equation}
where $\boldsymbol{S}_i$ is the spin of atom $i$, $J_{ij}$ is the exchange interaction between Ni atoms on sites $i$ and $j$, and $\boldsymbol{D}$ is the magnetic anisotropy energy (MAE). Three types of nonequivalent exchange interactions are considered, represented by $J_{1}$, $J_{2}$, and $J_{3}$, which correspond to first, second, and third NN exchange interactions, as illustrated in Fig.~\ref{nix2}(a).
The calculated exchange interactions are $J_{1}$ = -5.88 meV, $J_{2}$ = 0.19 meV, and $J_{3}$ = 3.18 meV.
The exchange interactions in the NiBr$_{2}$ monolayer satisfy the condition given by Eq.(\ref{eq:5}) in Ref.~\cite{ni}, indicating the presence of HM states.
\begin{equation}
	\begin{aligned}
	\frac{J_1 +3J_2 +4J_3 }{J_2-J_1+4J_3} > 0
	\end{aligned}\label{eq:5}
\end{equation}
As in the case of the NiI$_{2}$ monolayer, the frustration arising from the strong ferromagnetic (FM) interaction $J_{1}$ and the antiferromagnetic (AFM) interaction $J_{3}$ plays a key role in the formation of the HM state in the NiBr$_{2}$ monolayer.
Furthermore, the NiBr$_{2}$ monolayer exhibits an easy axis along the \(a\)-axis with a MAE of 0.20 meV, as calculated, which suggests that the NiBr$_{2}$ monolayer exhibits a cycloid spin order, as opposed to the proper-screw spin order in NiI$_{2}$~\cite{ni}. 



We further investigate the magnetic properties of the NiBr$_{2}$ monolayer using MC simulations. Based on the calculated exchange interactions, our simulations suggest that the ground state of the system is a noncollinear helical state with a Néel temperature \(T_{N}\) = 19.05 K, which closely matches the experimental results from Ref.~\cite{7,2021noncollinear}.  
In addition, we obtain six symmetrically equivalent modulation vectors \(\boldsymbol{Q}\) from the magnetic ground states, which are parallel to the [100], [010], and [110] directions in real space (as shown in Fig.~\ref{nix2}(b)).  
To distinguish between cycloidal and proper-screw orders, we perform direct DFT calculations, taking SOC into consideration. For simplicity, we focus on the case of \(\boldsymbol{Q}_1 = (0.2, -0.1, 0)\) in the [100] direction, which can be simulated using a 5$\times$10$\times$1 supercell for both cycloidal and proper-screw HM states.  
Our DFT calculations show that the cycloidal HM state is energetically more stable, with an energy 1.69 meV/Ni lower than that of the proper-screw state.  
We therefore propose that the magnetic ground state of the NiBr$_{2}$ monolayer is a cycloidal HM state, consistent with the spin-spiral calculations in Ref.~\cite{2021noncollinear}.  
Additionally, we find that the Br atoms develop a sizable magnetization of approximately 0.22 \(\mu_B\)/Br, forming a cycloidal HM state in coordination with that formed by the Ni spins (as reported in Ref.~\cite{ni}).

\subsection{Magnetoferroelectricity}

  We now turn our attention to magnetoferroelectric polarization. Electric polarization typically arises from two origins: ionic and electronic contributions, which are often comparable in multiferroic oxides~\cite{wu2012magnetic}. However, our DFT calculations show that lattice relaxations in the monolayer NiX$_{2}$ (X = I, Br) have a negligible impact on the electric polarization, indicating a dominant electronic origin. As a result, we exclude ionic contributions from our subsequent analysis of the electric polarization. The electric polarization amplitudes, \( |\boldsymbol{P}| \), are calculated to be approximately 1.30 \(\times\) 10$^{-13}$ C/m for monolayer NiI$_{2}$ and 2.67 \(\times\) 10$^{-13}$ C/m for monolayer NiBr$_{2}$, with directions perpendicular to their respective \(\boldsymbol{Q}\). The polarization vanishes when SOC is turned off, indicating that the magnetoelectric polarization is mediated by SOC. We further analyze the SOC-induced contributions to the polarization from each atomic type.

\begin{figure}[t]
\centering
	\includegraphics[scale=0.42]{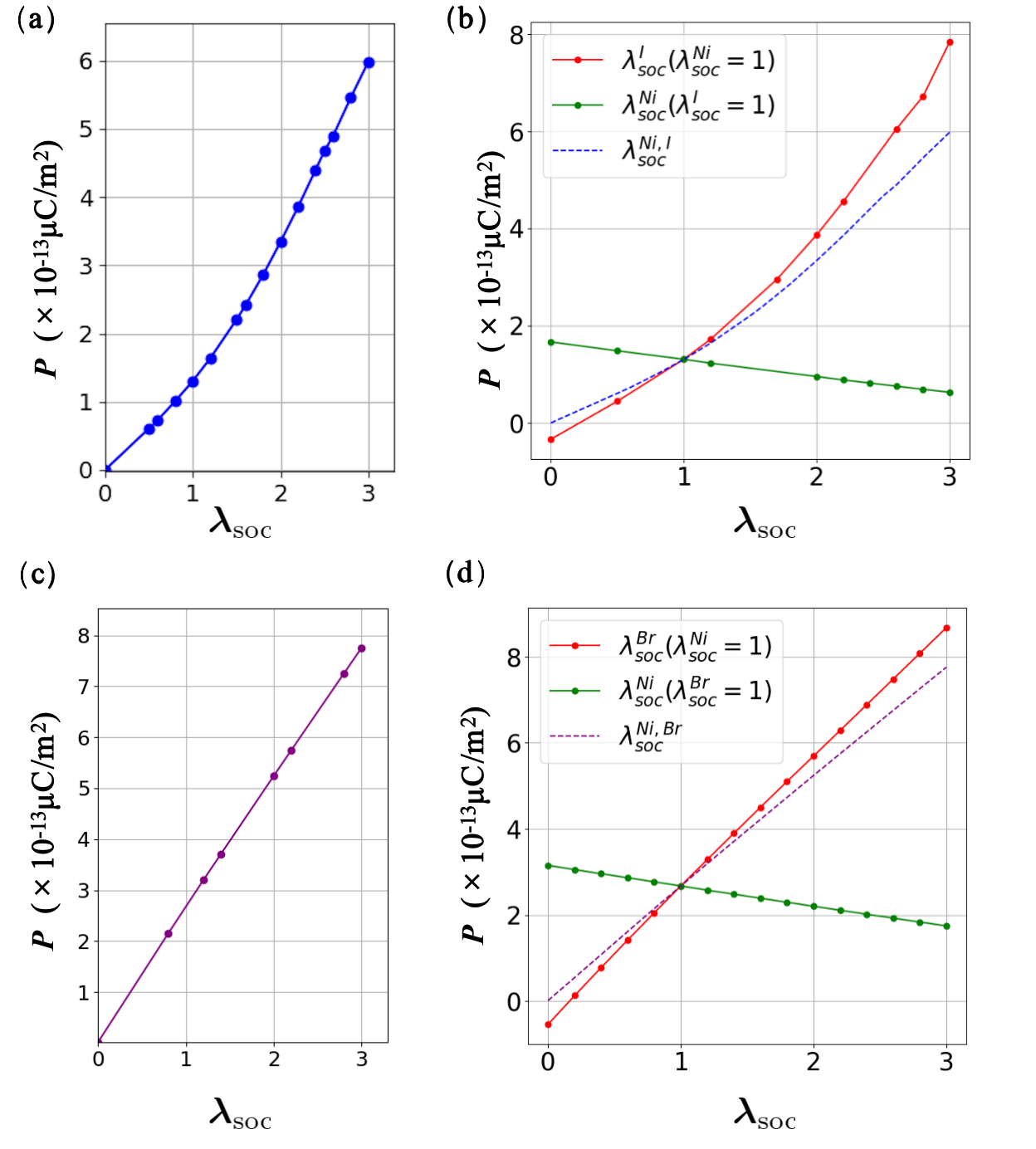}
	\caption{(a) and (c) The total ferroelectric polarization as a function of the dimensionless parameter \(\lambda_{\text{SOC}}\) for NiI\(_2\) (blue line) and NiBr\(_2\) (purple line), respectively. Panels (b) and (d) show the ferroelectric polarization as a function of the ligand SOC \(\lambda_{\text{SOC}}^{X}\), with \(\lambda_{\text{SOC}}^{Ni} = 1\) (red line), and as a function of the nickel SOC \(\lambda_{\text{SOC}}^{Ni}\), with \(\lambda_{\text{SOC}}^{X} = 1\) (green line).}
\label{tu2}
\end{figure}

We first assume that the total polarization $\boldsymbol{P}$ can be written as,

\begin{equation}
	\begin{aligned}
		\textbf{P} = \textbf{P}(\lambda_{SOC})
	\end{aligned}\label{eq:4}
\end{equation}
where \(\lambda_{SOC}\) denotes the strength of SOC, with \(\lambda_{SOC} = 1\) representing real SOC strength without artificial adjustment. We calculate the electric polarization by varying both \(\lambda_{SOC}^{X}\) and \(\lambda_{SOC}^{Ni}\) simultaneously. As shown in Fig.~\ref{tu2}(b) and Fig.~\ref{tu2}(c), our results indicate that the electric polarization, when both \(\lambda_{SOC}^{X}\) and \(\lambda_{SOC}^{Ni}\) are varied simultaneously, is very similar to that obtained by fixing \(\lambda_{SOC}^{Ni} = 1\). This suggests that the halogen SOC is the dominant contributor to the ferroelectric polarization. Moreover, the electric polarization increases as \(\lambda_{SOC}^{X}\) is enhanced. In contrast, the polarization decreases as \(\lambda_{SOC}^{Ni}\) is increased. Interestingly, for NiBr$_{2}$, the direct sum of the electric polarization calculated by considering only the SOC of Ni or Br is equal to the polarization calculated with full SOC, as it satisfies the following relation:
\begin{align}
\mathbf{P}(\lambda_{\mathrm{SOC}}^{\mathrm{Ni},X}) &= \mathbf{P}(\lambda_{\mathrm{SOC}}^{\mathrm{Ni}}(\lambda_{\mathrm{SOC}}^{X} = 1))  + \mathbf{P}(\lambda_{\mathrm{SOC}}^{X}(\lambda_{\mathrm{SOC}}^{\mathrm{Ni}} = 1)) \notag \\
&\quad - \mathbf{P}(\lambda_{\mathrm{SOC}}^{\mathrm{Ni},X} = 1)
\end{align}
However, this relation does not hold for NiI$_{2}$.

Next, we study the microscopic mechanism underlying the spin-induced ferroelectricity. Ref.~\cite{wang2016microscopic} classifies the polarization into intrasite \( \mathbf{P}_i (\mathbf{S}_i) \) and bilinear intersite \( \mathbf{P}_{ij}(\mathbf{S}_i \cdot \mathbf{S}_j) \) contributions, where \(i \neq j\). The intrasite polarization results from the variation in \(p\)-\(d\) hybridization due to SOC. Jia \textit{et al.}~\cite{jia2007microscopic}  have considered a cluster of two transition metal ions linked by a ligand ion, with the polarization contribution given by
\(
\mathbf{P} \propto (\mathbf{S}_i \cdot \mathbf{e}_{ij}) \mathbf{S}_i - (\mathbf{S}_j \cdot \mathbf{e}_{ij}) \mathbf{S}_j.
\)
This involves a single-spin mechanism, given by \( \mathbf{P} \propto \sum_i (\mathbf{S} \cdot \mathbf{e}_i)^2 \mathbf{e}_i \), where \( \mathbf{e}_i \) is unit vectors connecting the magnetic and ligand ions~\cite{yamauchi2011theoretical, murakawa2010ferroelectricity, jia2006bond}. However, for NiX$_{2}$, the intrasite term vanishes, as the Ni spins are located at inversion centers~\cite{wang2016microscopic}.

For intersite contributions, there are three widely recognized terms, the antisymmetric gKNB term~\cite{xiang2011general,zhang2017spin,zhang2017magnetic,xiang2013unified}, the anisotropic symmetric term~\cite{feng2016anisotropic}, and the exchange-striction term. Specifically, exchange striction is independent of SOC, while the other two mechanisms depend on SOC.
Since the polarization of the monolayer NiX$_2$ vanishes when SOC is turned off, the exchange-striction mechanism can be excluded. The Kitaev interaction, an anisotropic symmetric interaction, has been proposed to reproduce the experimentally observed in-plane propagation direction \(\langle 1\overline{1}0 \rangle\) in NiI$_2$, and was once considered the origin of ferroelectricity~\cite{li2023realistic}. However, the symmetric anisotropic polarization vanishes when an inversion center is located at the midpoint of the bond connecting \(\mathbf{S_{i}}\) and \(\mathbf{S_{j}}\) (see Appendix A for proof). Therefore, while the Kitaev interaction is considered to critically affect the magnetic ground state of NiI$_2$, it does not directly serve as the microscopic origin of the ferroelectric polarization.

\subsection{monolayer NiBr$_{2}$}

In the subsequent analysis, if \(\lambda_{SOC}\) is not explicitly specified, polarizations are, by default, calculated with real SOC strength \(\lambda_{SOC} = 1\).  We first turn to the gKNB model~\cite{xiang2011general} to interpret the microscopic origin of the magnetoelectric coupling in the NiBr$_{2}$ monolayer. According to the gKNB model, a magnetoelectric matrix \( M \) can be assigned to each Ni-Ni dimer~\cite{xiang2011general}. The polarization of the NiX$_{2}$ monolayer can then be expressed as
\begin{equation}
	\begin{aligned}
		\mathbf{P} = \sum_{ij} M^{ij} (\mathbf{S}_i \times \mathbf{S}_j),
	\end{aligned}\label{eq:2}
\end{equation}

where \( M^{ij} \) represents the magnetoelectric matrix between the spin dimers at sites \( i \) and \( j \), and \( \mathbf{S} \) denotes the normalized spin vectors. To begin, we select a first NN spin dimer aligned along the Cartesian \( x \)-axis and calculate the corresponding \( M \) matrix using the four-state method~\cite{xiang2011general}. The calculated \( M \) matrix takes the following form:

\begin{equation}
\begin{aligned}
M^{1\langle \text{x-axis} \rangle} =
\begin{pmatrix}
    M_{11} & 0 & 0 \\
    0 & M_{22} & M_{23} \\
    0 & M_{32} & M_{33}
\end{pmatrix}
\end{aligned}
\label{matrixform}
\end{equation}

Note that the superscript of the \( M \) matrix specifies the order of the NN and the orientation of the spin dimer. A comprehensive derivation of this matrix form is provided in Appendix A.

The \( M \) matrix elements in Eq.~\ref{matrixform} can be largely understood as contributions from two distinct spin-current mechanisms~\cite{miyahara2016theory,zhang2017spin,albaalbaky2017magnetoelectric,bolens2018theory}. The conventional spin-current mechanism, described by the KNB model, is expressed as \( \boldsymbol{P}_\text{KNB} \propto \boldsymbol{e}_{ij} \times (\boldsymbol{S}_i \times \boldsymbol{S}_j) \)~\cite{katsura2005spin}, where \( \boldsymbol{e}_{ij} \) is the unit vector connecting the spin pair. 
For specific high-symmetry bonds, the KNB model can impose the condition that \( M_{32} = -M_{23} = C \) are the only nonzero elements of \( M \). Building on symmetry considerations, Kaplan and Mahanti~\cite{kaplan2011canted} introduced an additional contribution, \( \boldsymbol{P}_\text{KM} \propto \boldsymbol{S}_i \times \boldsymbol{S}_j \), to account for spin-induced ferroelectricity in proper-screw helical magnetic structures. This mechanism results in nonzero diagonal elements of the \( M \) matrix.


The calculated \( M \) matrix is presented in Table~\ref{tab:matrixnibr2}. 
The polarization contributed by each site \( i \), denoted as \( \boldsymbol{P}_i^{\text{tot}} \), can be written as \( \boldsymbol{P}_i^{\text{tot}} = \sum_{k = 1}^6 \boldsymbol{P}_i^k \), since each Ni spin site \( i \) has six NN Ni spins. Considering only the first NN dimers in the cycloidal HM state of NiBr\(_2\) with the wavevector \(\boldsymbol{Q} = (2q, -q, 0)\), the total polarization \(\boldsymbol{P}_i^{\text{tot}}\) can be expressed as \(
\boldsymbol{P}_{i}^{\text{tot}} = \left(0, 2M^{1}_{23}(\sin4\pi q + \sin2\pi q), 2M^{1}_{33}(\sin4\pi q - 2\sin2\pi q)\right).
\)
This expression reveals that the total polarization \(\boldsymbol{P}_i^{\text{tot}}\) of NiBr\(_2\) is determined by the matrix elements \(M_{23}\) and \(M_{33}\). Notably, \(M_{33} = 0\), indicating that the polarization along the \(z\)-axis vanishes, consistent with symmetry analysis. The off-diagonal matrix element \(M_{23}\) represents the coupling coefficients arising from the \(\boldsymbol{P}_\text{KNB}\) mechanism.

However, the calculated magnitude of \(\left|\boldsymbol{P}(\text{M}^{1})\right|\) is $\sim$ \(1.47 \times 10^{-13} \, \text{C m}^{-1}\) (along $y$ direction), which is significantly smaller than the DFT result of \(\left|\boldsymbol{P}\right| \sim 2.67 \times 10^{-13} \, \text{C m}^{-1}\). We find that this discrepancy can be corrected by including contributions from the second and third NN dimers. The total polarization can then be expressed as
\begin{equation}
	\begin{aligned}
		\textbf{P} = \sum_{ij}M^{1}(\textbf{S}_i\ \times\ \textbf{S}_j)+\sum_{ij}M^{2}(\textbf{S}_i\ \times\ \textbf{S}_j)\\
		+\sum_{ij}M^{3}(\textbf{S}_i\ \times\ \textbf{S}_j)
	\end{aligned}\label{eq:3}
\end{equation}
where $M^1$, $M^2$ and $M^3$ represent the matrices for the first NN, second NN, and third NN dimers, respectively. 
The calculated values of \( M \) for monolayer NiBr$_2$ are listed in Table\ref{tab:matrixnibr2}.
The numerical results of Eq.~(\ref{eq:3}) are,
\begin{equation}
	\begin{aligned}
		\mathbf{P} & = \mathbf{P}({M}^{1}) +
		\mathbf{P}({M}^{2}) +
		\mathbf{P}({M}^{3}) \\
		& = (0.00, -1.47, 0.00) + 
		(0.00, -0.08, 0.00) + \\
		& \ \ \ (0.00, -1.21, 0.07) \times 10^{-13} \, \text{C m}^{-1} \\
		& = (0.00, -2.76, 0.07) \times 10^{-13} \, \text{C m}^{-1}
	\end{aligned}
	\label{eq:2br}
\end{equation}
The negligible $z$ component are from numerical errors. We therefore find that the gKNB model, considering the first, second, and third NN Ni-Ni dimers, reproduces the direct DFT results for the NiBr$_{2}$ monolayer quite well. The contribution of \( M^2 \) is found to be weak, while that of \( M^3 \) is comparable to that from \( M^1 \). In addition, the DFT polarizations are perfectly linear with respect to the \(\lambda_{SOC}\), aligning with the linear \(\lambda_{SOC}\) dependence of \(\boldsymbol{P}_\text{KNB}\)\cite{jia2007microscopic,kaplan2011canted,miyahara2016theory}.

\begin{table}[h]
\caption{The $M$ matrix of monolayer NiBr$_2$ from the first, second, and third NN spin pairs, corresponding to the spin pairs connected by $J_1$, $J_2$ and $J_3$, repectively.}
	\centering
        \renewcommand{\arraystretch}{1.5} 
 \setlength{\tabcolsep}{12pt} 

	\begin{tabular}{cccc}
    \hline
     \hline
    magnetoelectric matrix& \multicolumn{3}{c}{Units($10^{-5}$ eÅ)} \\
    \hline
     $M^{1\langle \text{x-axis} \rangle}$ & -3.5 & 0 & 0 \\
    & 0 & -20.13 & 35.25\\
    & 0 & -2.38 & 0 \\
    \hline
    $M^{2\langle \text{y-axis} \rangle}$& 0& 0 & -1.75\\
    & 5.5& 0& 0\\
    & 0 & 0& 0\\
    \hline
    $M^{3\langle \text{x-axis} \rangle}$& 0& 0 & 0 \\
    & 0 & -1.75& 29.0\\
    & 0 & -1.75& 1.38\\
    \hline
     \hline
\end{tabular}
	\label{tab:matrixnibr2}
\end{table}


The total polarization contributed by site i can then be written
$\boldsymbol{P}_{i}^{y}=2[M^{1}_{23}(\sin4\pi q+\sin2\pi q)+\sqrt{3}M^{2}_{13}\sin6\pi q+M^{3}_{23}(\sin8\pi q+\sin4\pi q)]$,
which can be simplified using a Taylor expansion (in units of $10^{-5}$ eÅ).
\begin{equation}
\begin{aligned}
\textbf{P}_{i}^{y} &= (12 M_{23}^1 + 12 \sqrt{3} M_{13}^2 + 24 M_{23}^3) \pi q \\
&\quad + (-24 M_{23}^1 - 72 \sqrt{3} M_{13}^2 - 192 M_{23}^3) \pi^3 q^3 + O(q)^4
\end{aligned}
\end{equation}
Therefore, in the small \( q \) region, a linear dependence of $\boldsymbol{P}^{y}$ on \( q \) is obtained, consistent with experimental results from Ref.~\cite{tokunaga2011multiferroicity}.





\subsection{monolayer NiI$_{2}$}

We first use the gKNB model to analyze the magnetoelectric polarization in monolayer NiI$_{2}$. The calculated $M^{1\langle \text{x-axis} \rangle}$ is given by
\begin{equation}
\begin{aligned}
M^{1\langle \text{x-axis} \rangle} =
\begin{pmatrix}
-20.13 & 0 & 0 \\
0 & -90.13 & 132.88 \\
0 & -5.88 & -1.75
\end{pmatrix}
\end{aligned}
\label{eq:matrix}
\end{equation}
For the proper-screw HM state of NiI$_2$ with \( \boldsymbol{Q} = (q, 0, 0) \), the total polarization \( \boldsymbol{P}_i^{\text{tot}} \) from the first-NN dimers at site \( i \) can be written as
\(
\boldsymbol{P}_i^{\text{tot}} = \left(\frac{\sqrt{3}}{2}A, -\frac{3}{2}A, 0 \right),
\)
where \( A = (M_{11}^{1} - M_{22}^{1}) \sin(2\pi q) \)~\cite{xiang2011general}.
Clearly, the total polarization \( \boldsymbol{P}_i^{\text{tot}} \) of NiI$_2$ is determined by the difference between the diagonal matrix elements \( M_{11} \) and \( M_{22} \), which reflects its origin in the mechanism \( \boldsymbol{P}_\text{KM} \).

\begin{figure}[t]  
	\centering 
	\includegraphics[scale=0.32]{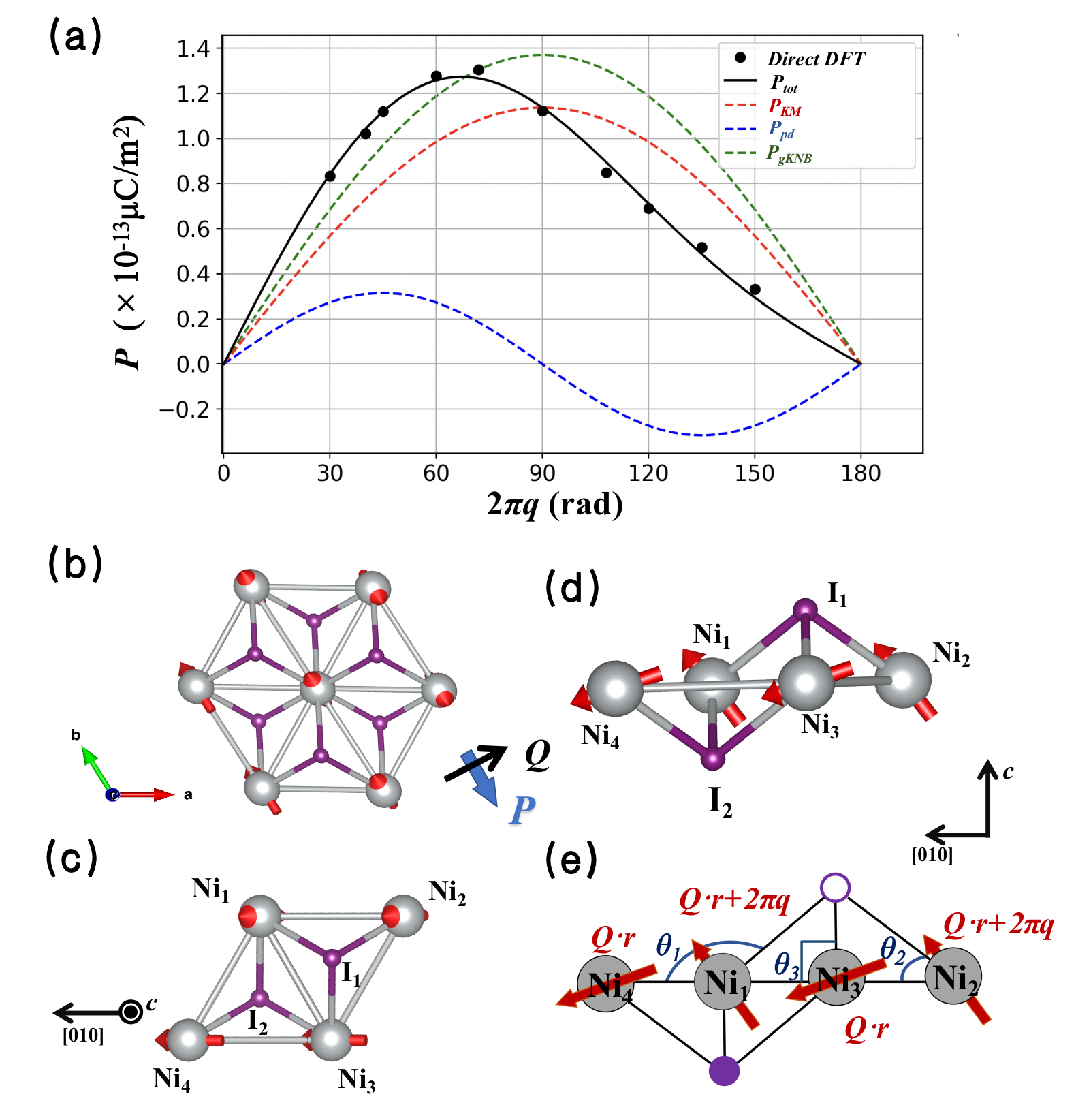}  
	\caption{(a) Black dots represent the total electric polarization from direct DFT calculations on the proper-screw magnetic structure with \( \boldsymbol{Q} = (q, 0, 0) \), together with a corresponding fitted curve (black lines). The dashed green line represents the calculated polarization based on the gKNB model. The red dashed green line and blue line denote the extracted polarization \( P_{\text{KM}} \) and \( P_{pd} \) respectively.
(b) Top view of a triangular lattice layer of Ni ions, along with two adjacent I layers. 
(c)-(d) Various perspectives of the Ni$_{2}$I$_{4}$ cluster used to calculate the electric polarization induced by the \( p \)-\( d \) hybridization mechanism.
(e) Local geometry and the phases of the spin directions on the Ni sites of the Ni$_{2}$I$_{4}$ cluster. The angle between the Ni$_{i}$-I$_{1}$ bond direction and the spin spiral plane (bc plane) is denoted as \( \theta_{i} \).}
	\label{pd}
\end{figure}

The polarization calculated from the first NN \( M \) is \( |\boldsymbol{P}(M^{1})| \sim 1.32 \times 10^{-13} \, \text{Cm}^{-1} \) (along the \([0\overline{1}0]\) direction), only slightly larger than the direct DFT result \( |\boldsymbol{P}| \sim 1.30 \times 10^{-13} \, \text{Cm}^{-1} \). Numerically, this suggests that the gKNB term for the first NN effectively explains the primary contribution to the polarization obtained from the direct DFT calculations for \(\boldsymbol{Q} = (0.2, 0, 0)\). As described by the gKNB formula, the total polarization is expected to be proportional to \( \sin2\pi q \). To further investigate the \( q \)-dependence of the polarization, we further calculate polarizations on magnetic configurations with \(\boldsymbol{Q} = (q, 0, 0)\) by varying \( q \) values. However, as shown in Fig.~\ref{pd}(a), the total polarization from direct DFT calculations deviates significantly from the calculated gKNB  polarization, suggesting that additional mechanisms are required to fully describe the polarization.



We further propose that the additional mechanism involves higher-order \( p \)-\( d \)-hybridization~\cite{arima2007ferroelectricity}. However, the electric polarization \( P_{\mathit{pd}} \propto \sin 4\pi q \) specifically applies to a proper-screw HM with \( \boldsymbol{Q} = (q, q, 0) \), as initially proposed by Arima~\cite{arima2007ferroelectricity}. Therefore, we extend Arima's model to monolayer NiI$_2$ for the case of a proper-screw HM with \( \boldsymbol{Q} = (q, 0, 0) \).




As illustrated in Fig.~\ref{pd}(c), the higher-order effect on the Ni$_{2}$I$_{4}$ cluster at a position \(\boldsymbol{r}\) is driven by the hybridization of one iodine ion with its three neighboring Ni$^{2+}$ ions. The phases of the spin directions at the Ni$_{1}$ (Ni$_{2}$) and Ni$_{3}$ (Ni$_{4}$) sites within the spin spiral plane (the $bc$ plane) are given by \(\boldsymbol{Q} \cdot \boldsymbol{r} + 2\pi q\) and \(\boldsymbol{Q} \cdot \boldsymbol{r}\), respectively, where \(\boldsymbol{Q}\) is the wave number (as shown in Fig.~\ref{pd}(e)). According to Arima's model, the charge transfer between the metal and ligand ions is influenced by the other two conjoint metal-ligand bonds. For example, if the covalency between Ni$_{2}$ and I$_{1}$ increases, the covalency of the Ni$_{3}$-I$_{1}$ and Ni$_{1}$-I$_{1}$ bonds weakens accordingly. Thus, the local polarization along the I$_{1}$-Ni$_{2}$ bond is modified as follows:
\begin{equation}
\small
\begin{aligned}
\mathbf{P}_{[I_1-Ni_2]}&=\left\{1 + \alpha \cos2 (\mathbf{Q} \cdot \mathbf{r} +2\pi q - \theta_2)\right\}\\
&\times\left\{1 - \beta \cos2 (\mathbf{Q} \cdot \mathbf{r} - \theta_3) - \gamma \cos2 (\mathbf{Q} \cdot \mathbf{r} +2\pi q - \theta_1)\right\} 
\end{aligned}
\end{equation}
where \(\alpha\), \(\beta\), and \(\gamma\) are constants. \(\theta_{i}\) represents the projection angle determined by the direction of the Ni\(_{i}\)-I\(_{1}\) bond within the spin spiral plane (the \(bc\)-plane), as illustrated in Fig.~\ref{pd}(e). Note that \(\beta\) and \(\gamma\) differ due to the tilting of the Ni\(_{3}\)-I\(_{1}\) and Ni\(_{1}\)-I\(_{1}\) bonds relative to the spin spiral plane. The geometric parameters \(\theta_{i}\) satisfy the conditions \(\theta_{3} = 90^\circ\) and \(\theta_{1} + \theta_{2} = \pi\).
The induced \(p\)-\(d\) polarization along the [010] direction, \(\boldsymbol{P}^{[010]}_{pd}\), is estimated by the imbalance in the charge transfer between Ni\(_{2}\)-I\(_{1}\) and Ni\(_{1}\)-I\(_{1}\) (here, $\mathbf{P} \perp \mathbf{Q}$). The uniform component induced by the \(p\)-\(d\) hybridization mechanism can then be expressed as
\begin{equation}
		\textbf{P}_{pd}^{[010]} = C\sin(\theta_{1}-\theta_{2})\sin4\pi q
        \label{eq:pd}
\end{equation}
where \( C \) is a constant.
For the proper-screw spin structure with \( \boldsymbol{Q} = (q, 0, 0) \), our derivation confirms that the magnitude of the \( p \)-\( d \) polarization keeps the form of \( P_{\mathit{pd}} \propto \sin(4 \pi q) \).

Taking $P_{\mathit{pd}}$ into account, the polarization can now be expressed in terms of its dependence on \( q \) as follows:
\begin{equation}
\boldsymbol{P}_{total}(q)=P_\text{KM}+P_\text{pd}=A\sin2\pi q+B\sin4\pi q
\label{poleq}
\end{equation}

By fitting Eq.~\ref{poleq} to the data from direct DFT calculations, we obtain \( A = 1.14 \times 10^{-13} \, \text{Cm}^{-1} \) and \( B = 3.15 \times 10^{-14} \, \text{Cm}^{-1} \). Specifically, for monolayer NiI\(_{2}\) with \( \boldsymbol{Q} = (0.2, 0, 0) \), the ratio of \( P_{\mathit{pd}} \) to \( P_{\text{KM}} \) is approximately \( 17\% \). As shown in Fig.~\ref{pd}(a), although \( P_{\mathit{pd}} \) is notably weaker than \( P_{\text{KM}} \), it plays a crucial role in capturing the correct \( q \)-dependence of the total polarization. We find that \( P_{\text{KM}} + P_{\mathit{pd}} \) effectively reproduces the total polarization in terms of the \( q \)-dependence.


We further investigate the \( \lambda_{\mathrm{SOC}} \) dependence of the polarization. Since \( P_{\text{KM}} \) is assumed to have a linear dependence on \( \lambda_{\mathrm{SOC}} \)~\cite{zhang2017spin, kaplan2011canted}, while \( P_{\mathit{pd}} \) is expected to exhibit a quadratic dependence\cite{arima2007ferroelectricity,kaplan2011canted,zhang2017spin}, the interplay between these two mechanisms governs the overall behavior of the total polarization, \( P_{\text{total}} \), as a function of \( \lambda_{\mathrm{SOC}} \). For \( \boldsymbol{Q} = (q, 0, 0) \), the total polarization can then be approximated as:

\begin{equation}
\small
\begin{aligned}
\boldsymbol{P}’_{total}(\lambda_{SOC})&=P_\text{KM}(\lambda_{SOC})+P_{pd}(\lambda_{SOC})\\
&=(A\sin2\pi q)\lambda_{SOC}+(B\sin 4\pi q)\lambda_{SOC}^2
\label{poleSOC}
\end{aligned}
\end{equation}

\begin{figure}[t]
    \centering
	\includegraphics[scale=0.25]{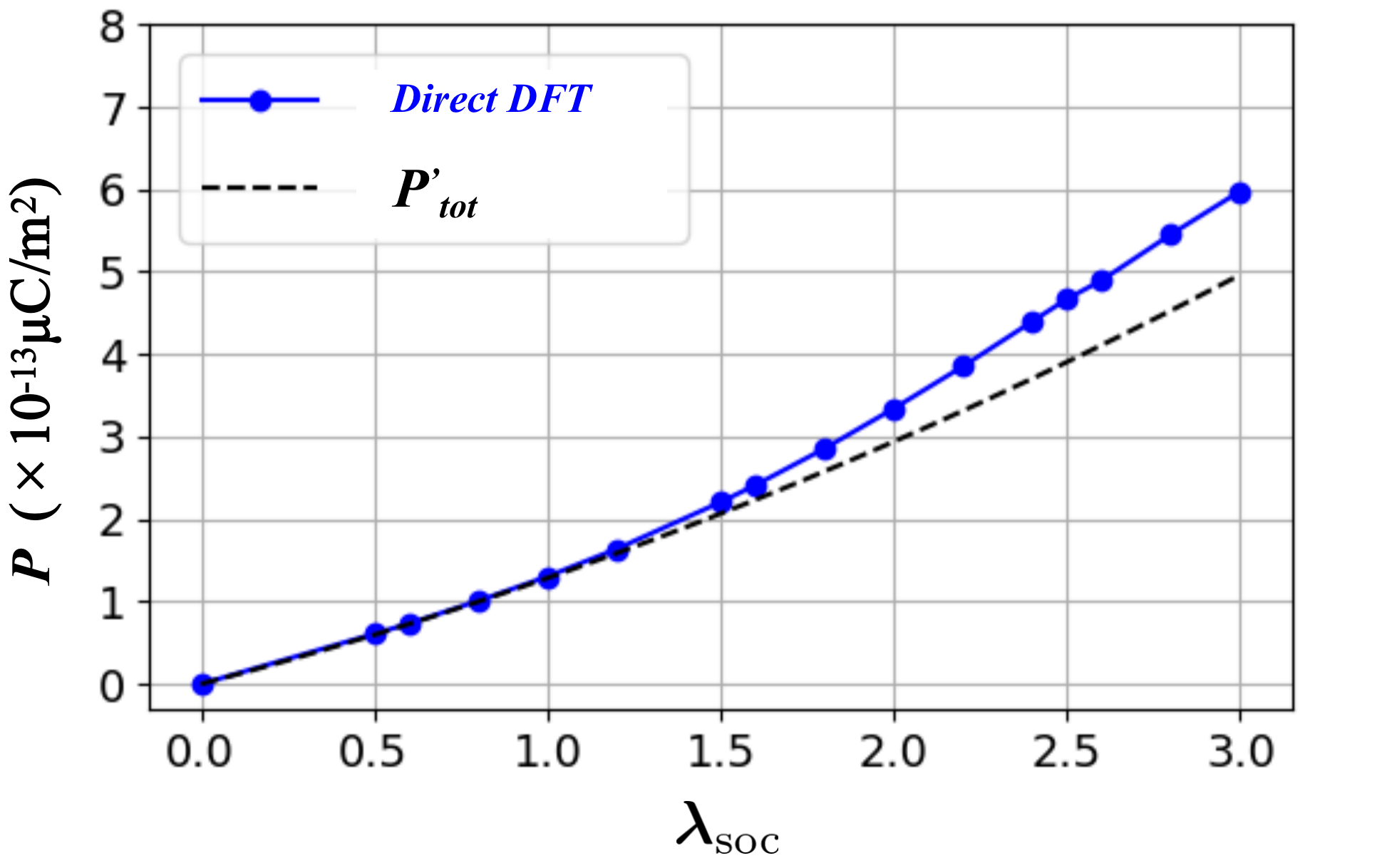}
	\caption{ The blue line represents the DFT calculated \( \boldsymbol{P} \) as a function of \( \lambda_{\text{SOC}} \)-dependence. The black line represents the \( \boldsymbol{P'}_{\text{tot}} \) in calculated by Eq.~\ref{poleSOC}. }
	\label{lambda-dependence}
\end{figure}

As shown in Fig.~\ref{lambda-dependence}, for \( \boldsymbol{Q} = (0.2, 0, 0) \), the overall \( \lambda_{\mathrm{SOC}} \) dependence predicted by Eq.~\ref{poleSOC} agrees well with the data from direct DFT calculations. Not surprisingly, the discrepancy increases with larger \( \lambda_{\mathrm{SOC}} \), where \( P_{\text{KM}}(\lambda_{\mathrm{SOC}}) \) and \( P_{\mathit{pd}}(\lambda_{\mathrm{SOC}}) \) are expected to deviate from their original linear and quadratic dependence.

The \( p \)-\( d \) hybridization mechanism is predicted to exist in various spin-spiral structures on a triangular lattice, as reported in Ref.~\cite{kurumaji2020spiral}. In contrast to monolayer NiI$_{2}$, the NiBr$_{2}$ monolayer exhibits a perfectly linear dependence of polarization on \( \lambda_{\text{SOC}} \), suggesting that the \( p \)-\( d \) hybridization mechanism can be largely excluded based on the \( \lambda_{\text{SOC}} \) dependence alone. Therefore, a quadratic-like \( \lambda_{\text{SOC}} \)-dependence of polarization may serve as indirect evidence for the presence of the \( p \)-\( d \) hybridization contribution.


\section{Summary}

In summary, we investigate the magnetoelectric properties of monolayer NiX$_2$ using first-principles calculations. We predict that the NiBr$_2$ monolayer stabilizes in a cycloidal magnetic ground state, with electric polarization linearly dependent on \(\lambda_{\text{SOC}}\), which can be well explained by the gKNB model. 
For the NiI$_2$ monolayer, a proper-screw helical magnetic ground state with the experimentally observed modulation vector is incorporated into our calculations.
However, the gKNB model alone cannot fully capture the complex dependence of electric polarization on both \(q\) and \(\lambda_{\text{SOC}}\) in NiI$_2$. To address this, we introduce the higher-order \(p\)-\(d\) hybridization mechanism, proposing that the magnetoelectricity in NiI$_2$ arises from the combined effects of the gKNB model and \(p\)-\(d\) hybridization. Our analysis quantitatively separates the contributions of these two mechanisms, based on the \(q\)- and \(\lambda_{\text{SOC}}\)-dependence of polarization. Overall, our study provides a comprehensive understanding of the microscopic origins of multiferroicity in NiBr$_2$ and NiI$_2$ monolayers, with insights applicable to their bulk counterparts.

\section{acknowledgments}
	
Work at Sun Yat-Sen University was supported by National Key Research and Development Program of China (Grants No. 2022YFA1402802, 2018YFA0306001), Guangdong Basic and Applied Basic Research Foundation (Grants No. 2022A1515011618), National Natural Science Foundation of China (Grant No. 92165204), Leading Talent Program of Guangdong Special Projects (201626003), Guangdong Provincial Key Laboratory of Magnetoelectric Physics and Devices (No. 2022B1212010008), Guangdong Provincial Quantum Science Strategic Initiative (GDZX2401010), and Fundamental Research Funds for the Central Universities, Sun Yat-sen University (Grants No. 23ptpy158). This work is supported in part by The Major Key Project of Peng Cheng Laboratory (PCL).

\section{APPENDIX A: Symmetry analysis}
The polarization component \( P^{\alpha} \) caused by a pair of spins \( \mathbf{S}_i \) and \( \mathbf{S}_j \) can be written as:
\[
P^{\alpha} = \begin{pmatrix} S_i^x & S_i^y & S_i^z \end{pmatrix} 
\begin{pmatrix} 
\lambda_{xx}^{\alpha} & \lambda_{xy}^{\alpha} & \lambda_{xz}^{\alpha} \\ 
\lambda_{yx}^{\alpha} & \lambda_{yy}^{\alpha} & \lambda_{yz}^{\alpha} \\ 
\lambda_{zx}^{\alpha} & \lambda_{zy}^{\alpha} & \lambda_{zz}^{\alpha} 
\end{pmatrix} 
\begin{pmatrix} S_j^x \\ S_j^y \\ S_j^z \end{pmatrix}
\]
Considering the $M$ matrix for a bond along the  \( x \)-axis direction, with an inversion center located at the midpoint of the bond, we apply inversion operation,
\[
\begin{pmatrix} P^x \\ P^y \\ P^z \end{pmatrix} \to \begin{pmatrix} -P^x \\ -P^y \\ -P^z \end{pmatrix};
\begin{pmatrix} S_i^x \\ S_i^y \\ S_i^z \end{pmatrix} \to \begin{pmatrix} S_j^x \\ S_j^y \\ S_j^z \end{pmatrix};
\begin{pmatrix} S_j^x \\ S_j^y \\ S_j^z \end{pmatrix} \to \begin{pmatrix} S_i^x \\ S_i^y \\ S_i^z \end{pmatrix}
\]
It is important to note that spins are axial vectors, therefore, the inversion symmetry operation simply interchanges \( \mathbf{S}_i \) and \( \mathbf{S}_j \). The $M$ matrix must satisfy the relation
\(P^{\alpha} = -\mathbf{S}_i M^{\alpha} \mathbf{S}_j = \mathbf{S}_j M^{\alpha} \mathbf{S}_i\)
which implies that \( (M^{\alpha})^{T} = -M^{\alpha} \).
In this case, \( \lambda_{ii}^{\alpha} = 0 \) and \( \lambda_{ij} ^{\alpha}= -\lambda_{ji}^{\alpha} \).
Hence, the $M$ matrix can be written as,
\[
M^{\alpha} = \begin{pmatrix} 
0 & \lambda_{xy}^{\alpha} & \lambda_{xz}^{\alpha} \\ 
-\lambda_{xy}^{\alpha} & 0 & \lambda_{yz}^{\alpha} \\ 
-\lambda_{xz}^{\alpha} & -\lambda_{yz}^{\alpha} & 0
\end{pmatrix} \]
It can be concluded that the anisotropic symmetric term is neglected when there is spatial inversion symmetry in the spin dimer. The contribution to the polarization from a spin pair originates from the gKNB term (as confirmed by DFT calculations). This contribution can be expressed as a function of \( \mathbf{S}_i \times \mathbf{S}_j \), implying that a change in helicity (the sign of \( q \)) should cause the polarization reversal.

For the first and third NN, the symmetry elements also include the \(yz\) mirror plane, which is perpendicular to the spin dimer and passes through the midpoint of the selected bond along the \(x\)-axis. This mirror plane imposes certain restrictions on the $M$ matrix form presented in the main text as Eq.~\eqref{eq:matrix}.

\section{APPENDIX B: Derivation of the Matrix Form for the Second-Nearest Neighbor Spin Dimer}

Based on the conclusions in Appendix A, we investigate the $M$ matrix form for the second NN spin dimer using symmetry analysis. With the bonding vector along the \( y \)-axis, the symmetry elements include both inversion symmetry and the \( yz \) mirror plane.

There is no interchange between \( \mathbf{S}_i \) and \( \mathbf{S}_j \) because the mirror plane contains the spin dimer. An improper rotation \( \mathbf{R} \) affects the spin (an axial vector) solely through its pure rotation component, \( -\mathbf{R} \). Since the \( yz \) mirror plane is an improper operation, we have \( \mathbf{S}_i' = -\mathbf{R}\mathbf{S}_i \) and \( \mathbf{S}_j' = -\mathbf{R}\mathbf{S}_j \). The transformations are as follows:

\[
\begin{pmatrix} P^x \\ P^y \\ P^z \end{pmatrix} \to \begin{pmatrix} -P^x \\ P^y \\ P^z \end{pmatrix};
\begin{pmatrix} S_i^x \\ S_i^y \\ S_i^z \end{pmatrix} \to \begin{pmatrix} S_i^x \\ -S_i^y \\ -S_i^z \end{pmatrix};
\begin{pmatrix} S_j^x \\ S_j^y \\ S_j^z \end{pmatrix} \to \begin{pmatrix} S_j^x \\ -S_j^y \\ -S_j^z \end{pmatrix}
\]

In this case, the matrix form must satisfy the condition 
\[
\mathbf{R}M \mathbf{R}^{T} = -M\implies
M = \begin{pmatrix} 
0 & M_{12} &  M_{13} \\ 
 M_{21} & 0 & 0\\ 
 M_{31} & 0 & 0
\end{pmatrix} \]

As for the other bonds, their matrix can be derived from the threefold rotational symmetry. Moreover, our results obtained from first-principles calculations on monolayer NiX$_{2}$ are in good agreement with the aforementioned symmetry analysis.

%

\end{document}